\documentclass{PoS}
\usepackage{epstopdf,multirow}

\title{IR suppression of the Coulomb gauge gluon propagator in SU(3) Yang-Mills theory}

\ShortTitle{IR suppression of the Coulomb gauge gluon propagator in SU(3) Yang-Mills theory}

\author{\speaker{Yoshiyuki Nakagawa}\\%
        Research Center for Nuclear Physics, Osaka University\\
        Ibarakisi, Osaka 567-0044, Japan\\
        E-mail: \email{nkgw@rcnp.osaka-u.ac.jp}}

\author{Atsushi Nakamura\\
        Research Institute for Information Science and Education, Hiroshima University\\
        Higashi-Hiroshima 739-8521, Japan\\
        E-mail: \email{nakamura@riise.hiroshima-u.ac.jp}}

\author{Takuya Saito\\
        Integrated Information Center, Kochi University\\
        Kochi, 780-8520, Japan\\
        E-mail: \email{tsaitou@kochi-u.ac.jp}}

\author{Hiroshi Toki\\
        Research Center for Nuclear Physics, Osaka University\\
        Ibarakisi, Osaka 567-0044, Japan\\
        E-mail: \email{toki@rcnp.osaka-u.ac.jp}}

\abstract{
We calculate the equal-time transverse gluon propagator
in Coulomb gauge QCD using a SU(3) quenched lattice
gauge simulation on large lattices, up to 11 [fm$^4$].
We find that the equal-time gluon propagator shows scaling violation;
namely, the data for different lattice spacings do not fall on
top of one curve.
This problem is cured by discarding data at large momenta,
which suffer from discretization errors.
In the infrared region, the transverse gluon propagator is strongly
suppressed and shows a turnover at about 500 [MeV].
Fitting the power law ansatz to the data at small momenta
predicts the vanishing gluon propagator at zero momentum,
indicating the confinement of gluons.
}

\FullConference{8th Conference Quark Confinement and the Hadron Spectrum \\
		 September 1-6 2008\\
		 Mainz, Germany}

\begin{document}

\section{Introduction}

Among several scenarios of color confinement proposed
since the discovery of QCD, Coulomb gauge QCD has recently
been received much attention along lattice QCD simulations
and a variational approach.
Coulomb gauge is a physical gauge in the sense that unphysical
degrees of freedom, such as longitudinal component of gluons,
are integrated out and the color Gauss' law is formally solved.
As a result, an instantaneous interaction shows up in the Hamiltonian
in Coulomb gauge QCD, which plays a central role in the confinement
mechanism in Coulomb gauge.
In the Gribov-Zwanziger scenario, the path integral is dominated
by the configurations near the Gribov horizon where the lowest
eigenvalue of the Faddeev-Popov (FP) operator vanishes, and
the eigenvalue distribution of the FP operator gets concentrated
near the vanishing eigenvalue compared to that in the abelian gauge theory
\cite{ZwanzigerD:NPB412:1994}.
Such an enhancement has been observed by the lattice simulations
\cite{GreensiteJ:JHEP05:2005,NakagawaY:PRD75:2007}.
Accordingly, the color-Coulomb instantaneous interaction is strongly
enhanced and provides a confining force between color charges.
Lattice QCD simulations have showed that the instantaneous
color-Coulomb potential rises linearly at large distances
and it is stronger than the static potential
\cite{GreensiteJ:PRD67:2003,NakamuraA:PTP115:2006,
NakagawaY:PRD73:2006,NakagawaY:PRD77:2008}.
This is an expected result from the Zwanziger's inequality
\cite{ZwanzigerD:PRL90:2003}.
On the other hand, the color-Coulomb propagator has been
evaluated by inverting the FP matrix, and it has been shown that
the color-Coulomb string tension almost saturates
the Wilson string tension
\cite{VoigtA:PRD78:2008}.

In the Gribov-Zwanziger scenario, the would-be physical
gluon propagator is expected to be suppressed in the infrared (IR)
region due to the proximity of the Gribov horizon in the IR direction
\cite{ZwanzigerD:NPB364:1991}.
In this study, we calculate the equal-time transverse gluon propagator,
\begin{equation}
D^{ab}_{\mu\nu}(\vec{x}-\vec{y})
= \langle A^a_{\mu}(\vec{x})A^b_{\nu}(\vec{y})\rangle
= D^{ab}_{\mu\nu}(\vec{x}-\vec{y}),
\end{equation}
in the momentum space,
\begin{equation}
D^{ab}_{ij}(\vec{p})
=\delta^{ab}\left(\delta_{ij}-\frac{p_ip_j}{\vec{p\,}^2}\right)
D^{\mathrm{tr}}(|\vec{p}|).
\end{equation}
using SU(3) quenched lattice QCD simulations.
The lattice configurations are generated by the heat-bath
Monte Carlo technique with the Wilson plaquette action.
In these simulations we adopt the iterative method with
the Fourier acceleration to fix a gauge, and the gauge fixing
is stopped if $(\partial_iA_i)^2<10^{-14}$ at each time slice.
The details of the simulations will be published elsewhere.

\section{Equal-time transverse gluon propagator}

\begin{figure}[htbp]
\begin{minipage}{0.47\hsize}\begin{center}
\resizebox{1.\textwidth}{!}
{\includegraphics{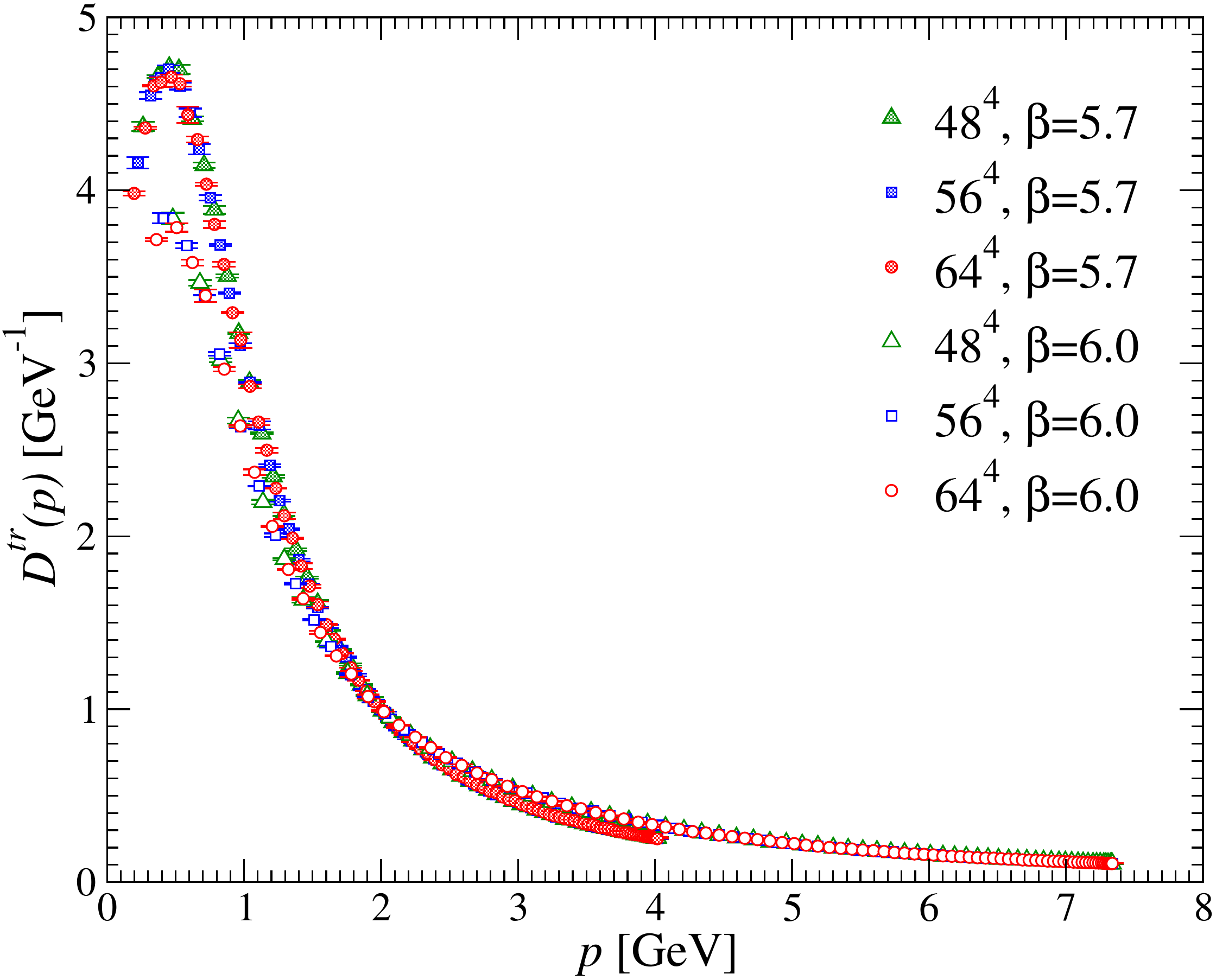}}
\end{center}\end{minipage}
\hspace{0.03\hsize}
\begin{minipage}{0.47\hsize}\begin{center}
\resizebox{1.\textwidth}{!}
{\includegraphics{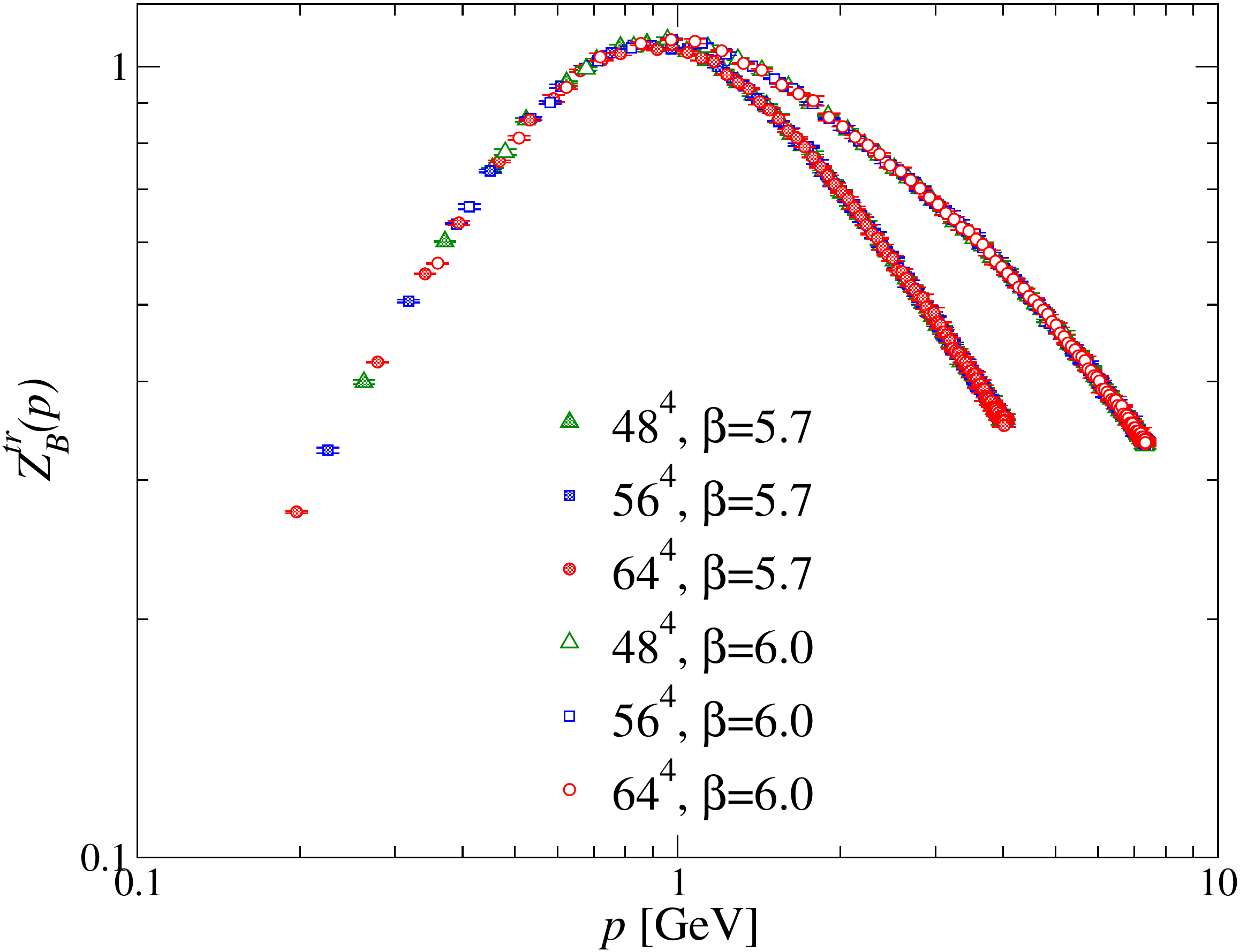}}
\end{center}\end{minipage}
\caption{
(Left) The equal-time transverse gluon propagator in physical units
at $\beta=5.7$ and $\beta=6.0$.
Data are converted from lattice units to physical units
by adopting the Necco-Sommer scaling relation.
The propagator is renormalized at 2 [GeV].
(Right) The dressing function of the unrenormalized
equal-time transverse gluon propagator,
$Z^{tr}_B(\vec{p}) = |\vec{p}|D^{tr}_B(\vec{p})$,
at different lattice couplings is plotted in physical units.
}
\label{Dij_scaling_violation}
\end{figure}

The equal-time transverse gluon propagator
at $\beta$=5.7 and 6.0 is drawn
in the left panel of Fig. \ref{Dij_scaling_violation}.
The cone cut and the further cut are applied and
the propagator is normalized such that
$D^{tr}(|\vec{p}|=2\textrm{ [GeV]}) = 1$.
The scale is set by using the Necco-Sommer scaling relation
\cite{NeccoS:NPB622:2002}.
We observe that the gluon propagator has a maximum
at $p = 0.4 \sim 0.5$ [GeV] irrespective of the lattice coupling
and it decreases with decreasing the momentum in the IR region.
This is the striking feature of the gluon propagator.
The equal-time propagator is defined as the energy integral
of the 4-dimensional propagator,
\begin{equation}
D^{eq}(|\vec{p}|) = \int \frac{dp_4}{2\pi} D(\vec{p}, p_4),
\end{equation}
and it can be interpreted as the inverse of
the energy dispersion relation.
Thus, the propagator at vanishing momentum corresponds to
the inverse of the effective mass of the gluon.
The IR suppression of the equal-time transverse gluon propagator
implies that the gluons have momentum dependent effective mass
$M(\vec{p})$, and it diverges in the IR limit,
$\lim_{\vec{p} \to 0} M(\vec{p})=\infty$,
indicating the confinement of gluons.

The left panel of Fig. \ref{Dij_scaling_violation}
also illustrates that the equal-time transverse gluon propagator
shows scaling violation; namely, the data points at different lattice
couplings cross at the renormalization point $|\vec{p}|=2$ [GeV]
and do not not fall on top of the same curve
\cite{NakagawaY:PSLAT2007:2007,VoigtA:PSLAT2007:2007,
BurgioG:2008}.
Taking a closer look at the raw results of the numerical simulations
gives us a clue to cure scaling violation.
The right panel of Fig. \ref{Dij_scaling_violation} shows
the dressing function of the unrenormalized equal-time gluon propagator
in physical units.
We observe that the dressing function at different lattice couplings
shows completely different behavior at high momenta.
On the other hand, the momentum dependence of the dressing
function is quite similar in the small momentum region.
This implies that the scaling violation can be cured by restricting
the momentum to
\begin{equation}
|p_{\mu}a| \le \alpha < 2,
\end{equation}
and discarding data at large momenta
where data suffer from strong discretization errors.
In order to find a reasonable value for $\alpha$ which
guarantees the scaling behavior of the transverse gluon propagator,
we adopt a matching procedure described in
\cite{LeinweberDB:PRD60:1999}.

\section{Result of the matching analysis and the IR behavior}

\begin{table}[htdp]
\begin{center}\begin{tabular}{cccc}
\hline\hline
$\alpha$ & $R_a$ & $R_Z$ & $\chi^2/ndf$ \\
\hline
0.7 & 1.73$^{1}_{2}$ & 0.967$^{3}_{4}$ & 9.72 \\
0.6 & 1.74$^{2}_{2}$ & 0.973$^{5}_{5}$ & 2.16 \\
0.5 & 1.74$^{2}_{2}$ & 0.974$^{6}_{7}$ & 2.72 \\
0.4 & 1.74$^{2}_{2}$ & 0.973$^{10}_{10}$ & 2.11 \\
\hline\hline
\end{tabular}\end{center}
\caption{
The matching results with the further cut $|p_ia|\le\alpha$.
$R_a=a(\beta=5.7)/a(\beta=6.0)$ is the ratio of the lattice spacings
and $R_Z=Z(\beta=5.7)/Z(\beta=6.0)$ is that of the renormalization
constants.
The Necco-Sommer scaling relation gives $R_a^{NS} = 1.83$.
}
\label{tab:match_Dij}
\end{table}

\begin{figure}[htbp]\begin{center}
\resizebox{0.5\textwidth}{!}
{\includegraphics{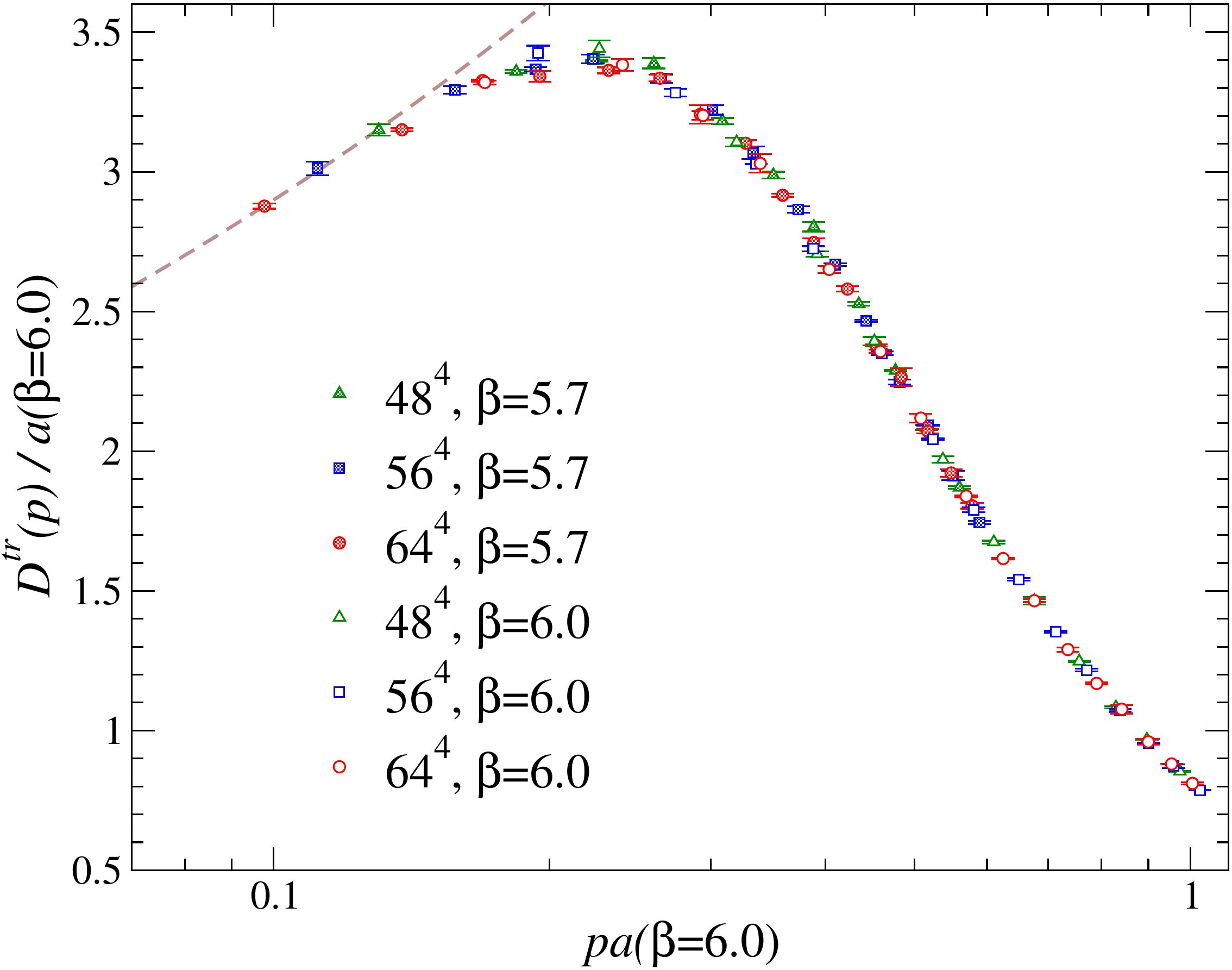}}
\caption{The matched gluon propagator with the further cut
$|p_ia| \le 0.6$.
The three momentum and the propagator are expressed
in units of the lattice spacing at $\beta=6.0$.
The dashed curve represents the fitting function
corresponding to the fitting range
$0 \le |\vec{p}_{max}|a(\beta=6.0) \le 0.135$.
}
\label{Dij_matched_060}
\end{center}\end{figure}

We cut data at large momenta and apply the matching procedure
to the transverse gluon propagator.
We refer to
\cite{LeinweberDB:PRD60:1999,NakagawaY}
for the details of the matching analysis.
We performed the matching between data at $\beta=6.0$
and $\beta=5.7$.
The ratios of the lattice spacing and the renormalization
constant obtained by the matching are given in Table \ref{tab:match_Dij}.
Although $\chi^2/ndf$ is relatively large for $\alpha = 0.7$,
it takes smaller values for smaller $\alpha$.
For $\alpha \le 0.6$, $\chi^2/ndf$ takes acceptable values,
and $R_a$ and $R_Z$ are stable against the change of $\alpha$.
The equal-time transverse gluon propagator with the further cut
$\alpha \le 0.6$ is shown in Fig. \ref{Dij_matched_060}.
We observe that the data points for different lattice couplings
nicely fall on the same curve.
This implies that scaling behavior of the equal-time gluon propagator
is recovered by cutting the data points at large momenta
which suffer from the discretization effects.

In order to explore the IR behavior of the equal-time transverse
gluon propagator, we make the power law ansatz,
\begin{equation}
D^{tr}(|\vec{p}|) = d_1 |\vec{p}|^{\gamma_{gl}^{IR}},
\end{equation}
in the IR region.
The fitted parameters are listed in Table \ref{tab:glue_IR}.
The fitting function corresponding to the smallest fitting range
is plotted in Fig. \ref{Dij_matched_060}.
Although the fitting becomes worse and the IR exponent
$\gamma^{IR}_{gl}$ gets small as the maximum momentum
of the fitting range increases,
$\gamma^{IR}_{gl}$ takes positive value in all cases.
That is, our result of the IR fitting predicts the vanishing transverse
gluon propagator at zero momentum.

\begin{table}[htdp]
\begin{center}\begin{tabular}{cccc}
\hline
$|\vec{p}_{max}|a(\beta=6.0)$ & $d_1$ & $\gamma_{gl}^{IR}$ & $\chi^2/ndf$ \\
\hline
0.135 & 6.02(32) & 0.318(24) & 0.20 \\
0.140 & 5.29(11) & 0.261(10) & 3.49 \\
0.160 & 5.38(10) & 0.269(9) & 3.33 \\
0.170 & 5.29(5) & 0.261(5) & 2.64 \\
\hline
\end{tabular}\end{center}
\caption{
The result of the IR power law fitting.
$|\vec{p}_{max}|a(\beta=6.0)$ represents
the maximum momentum of the fitting range.}
\label{tab:glue_IR}
\end{table}

\section{Summary and conclusion}

We calculate the equal-time transverse gluon propagator
on large lattices, up to 11 [fm$^4$].
We find that the equal-time gluon propagator shows scaling violation.
This problem is cured by discarding data at large momenta
which suffer from discretization errors.
In the IR region, the transverse gluon propagator is strongly
suppressed and shows the turnover at about 500 [MeV].
Fitting the power law ansatz to the data at small momenta
predicts the vanishing gluon propagator at zero momentum,
indicating the confinement of gluons.

\section*{Acknowledgements}

The simulation was performed on
NEC SX-8R at RCNP, Osaka University
and NEC SX-9 at CMC, Osaka University.
We appreciate the warm hospitality and support of the RCNP administrators.

\bibliographystyle{h-physrev4}

\begin{thebibliography}{9}

\bibitem{ZwanzigerD:NPB412:1994}
D.~Zwanziger,
\newblock Nucl. Phys. {\bf B412}, 657 (1994).

\bibitem{GreensiteJ:JHEP05:2005}
J.~Greensite, S.~Olejnik and D.~Zwanziger,
\newblock JHEP {\bf 05}, 070 (2005).

\bibitem{NakagawaY:PRD75:2007}
Y.~Nakagawa, A.~Nakamura, T.~Saito and H.~Toki,
\newblock Phys. Rev. {\bf D75}, 014508 (2007).

\bibitem{GreensiteJ:PRD67:2003}
J.~Greensite and S.~Olejnik,
\newblock Phys. Rev. {\bf D67}, 094503 (2003).

\bibitem{NakamuraA:PTP115:2006}
A.~Nakamura and T.~Saito,
\newblock Prog. Theor. Phys. {\bf 115}, 189 (2006).

\bibitem{NakagawaY:PRD73:2006}
Y.~Nakagawa, A.~Nakamura, T.~Saito, H.~Toki and D.~Zwanziger,
\newblock Phys. Rev. {\bf D73}, 094504 (2006).

\bibitem{NakagawaY:PRD77:2008}
Y.~Nakagawa, A.~Nakamura, T.~Saito and H.~Toki,
\newblock Phys. Rev. {\bf D77}, 034015 (2008).

\bibitem{ZwanzigerD:PRL90:2003}
D.~Zwanziger,
\newblock Phys. Rev. Lett. {\bf 90}, 102001 (2003).

\bibitem{VoigtA:PRD78:2008}
A.~Voigt, E.~M. Ilgenfritz, M.~M\"uller-Preussker and A.~Sternbeck,
\newblock Phys. Rev. {\bf D78}, 014501 (2008).

\bibitem{ZwanzigerD:NPB364:1991}
D.~Zwanziger,
\newblock Nucl. Phys. {\bf B364}, 127 (1991).

\bibitem{NeccoS:NPB622:2002}
S.~Necco and R.~Sommer,
\newblock Nucl. Phys. {\bf B622}, 328 (2002).

\bibitem{NakagawaY:PSLAT2007:2007}
Y.~Nakagawa, A.~Nakamura, T.~Saito and H.~Toki,
\newblock PoS {\bf LAT2007}, 319 (2007).

\bibitem{VoigtA:PSLAT2007:2007}
A.~Voigt, E.-M. Ilgenfritz, M.~M\"uller-Preussker and A.~Sternbeck,
\newblock PoS {\bf LAT2007}, 338 (2007).

\bibitem{BurgioG:2008}
G.~Burgio, M.~Quandt and H.~Reinhardt,
\newblock 0807.3291.

\bibitem{LeinweberDB:PRD60:1999}
D.~B. Leinweber, J.~I. Skullerud, A.~G. Williams and C.~Parrinello,
\newblock Phys. Rev. D {\bf 60}, 094507 (1999).

\bibitem{NakagawaY}
Y.~Nakagawa, A.~Voigt, E.-M. Ilgenfritz, M.~M\"uller-Preussker,
A.~Nakamura, T.~Saito, A.~Sternbeck, and H.~Toki,
\newblock in preparation.

\end{thebibliography}


\end{document}